\documentclass[11pt]{article}

\usepackage[left=2.5cm, right=2.5cm, top=2.5cm, bottom=2.5cm]{geometry}
\usepackage{amssymb,amsthm,amsmath,commath,mathtools}
\usepackage{braket}
\usepackage{xcolor}
\usepackage{algorithm}
\usepackage{graphicx}
\usepackage[noend]{algpseudocode}
\usepackage{authblk}
\usepackage{subfigure}
\usepackage{pgfplots}
\pgfplotsset{colormap/viridis}
\pgfplotsset{compat=1.18}

\usepackage[hidelinks, final=true,colorlinks = false,allcolors = {blue}]{hyperref}
\usepackage[capitalize]{cleveref}

\newtheorem{theorem}{Theorem}
\newtheorem{proposition}{Proposition}
\newtheorem{lemma}{Lemma}
\newtheorem{corollary}{Corollary}

\newtheorem{remark}{Remark}

\DeclareMathOperator{\Jac}{\mathrm{Jac}}

\newcommand{\real}{\mathbb{R}}
\newcommand{\integer}{\mathbb{Z}}

\usepackage{stmaryrd}
\usepackage{dsfont}

\DeclareFontFamily{U}{txsyc}{}
\DeclareFontShape{U}{txsyc}{m}{n}{
   <-> txsyc%
}{}
\DeclareFontShape{U}{txsyc}{bx}{n}{
   <-> txbsyc%
}{}
\DeclareFontShape{U}{txsyc}{l}{n}{<->ssub * txsyc/m/n}{}
\DeclareFontShape{U}{txsyc}{b}{n}{<->ssub * txsyc/bx/n}{}
\DeclareSymbolFont{symbolsC}{U}{txsyc}{m}{n}
\SetSymbolFont{symbolsC}{bold}{U}{txsyc}{bx}{n}
\DeclareFontSubstitution{U}{txsyc}{m}{n}
\DeclareMathSymbol{\df}{\mathrel}{symbolsC}{"42}
\newcommand{\me}{\medskip}
\newcommand{\sm}{\smallskip}

\newcommand{\st}{\,:\,}
\newcommand{\ri}{\rightarrow}
\newcommand{\iy}{\infty}
\newcommand{\ZZ}{\mathbb{Z}}
\newcommand{\RR}{\mathbb{R}}
\newcommand{\NN}{\mathbb{N}}
\newcommand{\LL}{\mathbb{L}}
\newcommand{\bq}{\begin{eqnarray*}}
\newcommand{\bqn}[1]{\begin{eqnarray}\label{#1}}
\newcommand{\eq}{\end{eqnarray*}}
\newcommand{\eqn}{\end{eqnarray}}
\newcommand{\lt}{\left}
\newcommand{\f}[2]{\frac{#1}{#2}}
\newcommand{\rt}{\right}
\newcommand{\lin}{\llbracket}
\newcommand{\rin}{\rrbracket}
\newcommand{\fo}{\forall\ }
\newcommand{\wwtbp}{\par\hfill $\blacksquare$\par\me\noindent}
\newcommand{\lve}{\lt\vert}
\newcommand{\rve}{\rt\vert}
\newcommand{\pa}{\partial}
\newcommand{\un}{\mathds{1}}

\newcommand{\dete}{\mathrm{det}}
\newcommand{\cG}{\mathcal{G}}

\newcommand{\tr}{\triangle}

\renewcommand{\epsilon}{\varepsilon}

\begin{document}

\allowdisplaybreaks

\title{Quantum walks, the discrete wave equation\\ and Chebyshev polynomials}
\author{Simon Apers\thanks{Universit\'e Paris Cit\'e, CNRS, IRIF, Paris, France. Email: smgapers@gmail.com} \, and Laurent Miclo\thanks{Toulouse School of Economics,
Institut de Mathématiques de Toulouse, CNRS and University of Toulouse. Email: miclo.math.cnrs.fr@gmail.com}}

\maketitle

\begin{abstract}
A quantum walk is the quantum analogue of a random walk.
While it is relatively well understood how quantum walks can speed up random walk hitting times, it is a long-standing open question to what extent quantum walks can speed up the spreading or mixing rate of random walks on graphs.
In this expository paper, inspired by a blog post by Terence Tao, we describe a particular perspective on this question that derives quantum walks from the discrete wave equation on graphs.
This yields a description of the quantum walk dynamics as simply applying a Chebyshev polynomial to the random walk transition matrix.
This perspective decouples the problem from its quantum origin, and highlights connections to earlier (non-quantum) work and the use of Chebyshev polynomials in random walk theory as in the Varopoulos-Carne bound.
We illustrate the approach by proving a weak limit of the quantum walk dynamics on the lattice.
This gives a different proof of the quadratically improved spreading behavior of quantum walks on lattices.
\end{abstract}

\newpage

\section{Introduction} \label{sec:intro}

Random walks (and Markov chains, more generally) show up in many areas of science.
A particularly relevant question is how fast random walks spread on graphs.
This is related to topics such as the thermodynamic equilibration of physical systems, the time complexity of Markov chain Monte Carlo algorithms, or the diffusion speed of diseases and rumours in networks.
It is typically quantified by the random walk \emph{mixing time} \cite{levin2017markov}, which captures the time that it takes for a random walk to converge to its stationary distribution.

Quantum walks are the quantum mechanical analogues of random walks.
They are a key tool in the design of quantum algorithms \cite{ambainis2003quantum}, and they have been a source of interesting mathematics.
Building on a long line of work, it was recently established \cite{ambainis2019quadratic,piddock2019quantum} that quantum walks can quadratically speed up the random walk \emph{hitting time}, i.e., the expected time to hit a subset of vertices when starting from the stationary distribution.
This can be thought of as a graph-theoretic version of the quadratic speedup in Grover's quantum search algorithm \cite{grover1996fast}.
The random walk mixing time is somehow dual to the hitting time, requiring instead to converge to the stationary distribution starting from a single state.
A significantly smaller body of work has been devoted to proving a quadratic quantum walk speedup in the mixing time, paralleling the quadratic speedup in the hitting time.
Such a speedup was proven in a number of relevant but very restricted settings, including settings where the underlying graph is highly symmetric \cite{aharonov2001quantum,richter2007almost,richter2007quantum}, very poorly connected \cite{apers2019qsampling}, or when we have access to a simulated annealing schedule \cite{wocjan2008speedup,harrow2020adaptive}.

Despite advances in these restricted settings, it seems that progress on a general quantum walk speedup in the random walk spreading rate or mixing time has largely stalled.
We revisit this question using an alternative perspective on the quantum walk dynamics, based on Chebyshev polynomials.\footnote{The connection between quantum walks and Chebyshev polynomials has been used before (see e.g.~\cite{childs2017quantum,gilyen2019quantum}) but not in the context of quantum walk mixing.}
Rather than following the standard exposition that is usually found in the quantum algorithms literature (see e.g.~\cite{childs2017lecture}), we present an alternative perspective based on the discrete wave equation on graphs.
This approach is inspired and largely follows a blog post on the discrete wave equation by Terence Tao \cite{taoblog} (which, on its turn, was inspired by discussions with Yuval Peres).
It also connects to earlier work on the discrete wave equation on graphs by Friedman and Tillich \cite{friedman2004wave}.
Surprisingly, the connection with quantum walks seems to not have been made earlier, and formalizing this is one of the motivations for writing this paper.
As the main technical contribution and proof-of-concept, we use the analysis of Chebyshev polynomials to prove a weak limit on the quantum walk spreading behavior.
In particular, this implies that quantum walks have a ballistic spreading rate on lattices, in correspondence with earlier findings \cite{mackay2002quantum,baryshnikov2011two} that used very different techniques.

Throughout the manuscript, we put care in giving a self-contained exposition, with the hope of rejuvenating interest in the problem and attracting a broader (and potentially non-quantum) audience.

\section{Random walks from the diffusion equation} \label{sec:RW}

As a warm-up, consider the random walk dynamics
\begin{equation} \label{eq:RW}
\fo t\in\ZZ_+,\qquad u(t+1)
= P u(t)
\end{equation}
where $P$ is the random walk transition matrix on a graph.
We will argue how these dynamics correspond to a discrete diffusion equation on graphs.

To this end, consider the diffusion equation
\bqn{diff}
\dot{u}
&=&\f1{2d} \Delta u
\eqn
over $\real^d$ with $\Delta$ the Laplacian operator.
A discrete space and time analogue of~\eqref{diff} is given by~\eqref{eq:RW} with $P$ the random walk transition matrix on the lattice $\integer^d$. 
Indeed, one gets an approximation of~\eqref{diff} by introducing a small parameter $\epsilon>0$ and replacing the time set $\ZZ$ by $\epsilon\ZZ$ and the lattice $\ZZ^d$ by $\sqrt{\epsilon}\ZZ^d$. Consider then the evolution
\bqn{ue}
\fo t\in\epsilon\ZZ_+,\qquad u_\epsilon(t+\epsilon)&=& P_{\sqrt{\epsilon}}[u_\epsilon(t)]
\eqn
where $P_{\sqrt{\epsilon}}$ is the random walk transition matrix on $\sqrt{\epsilon}\integer^d$.
This evolution coincides with \eqref{eq:RW} for~$\epsilon=1$.
Now assume e.g.\ that the initial condition $u(0)$ of~\eqref{diff} is continuous and has a compact support and that the initial condition of~\eqref{ue} is just the restriction of $u(0)$ to $\sqrt{\epsilon}\ZZ^d$.
Then the following well-known approximation result holds, as a consequence of Donsker's invariance principle \cite{donsker1952justification,morters2010brownian}: for any given $(t,x)\in\RR_+\times\RR^d$, we have
\bq
\lim_{\epsilon\ri0_+} u_\epsilon(t_\epsilon,x_\epsilon)&=&u(t,x)\eq
where for any $\epsilon>0$, $t_\epsilon\in\epsilon\ZZ_+$ and $x_\epsilon\in \sqrt{\epsilon}\ZZ^d$ are such that $\lim_{\epsilon\ri0_+}t_\epsilon = t$ and $\lim_{\epsilon\ri0_+}x_\epsilon = x$.

This motivates identifying a random walk on the lattice $\integer^d$ with a discrete diffusion equation.
Now, through a small leap of faith, we can generalize this by letting $P$ be the random walk transition matrix on a general graph.
We identify the corresponding dynamics in \eqref{eq:RW} with the discrete wave equation on that graph.

\section{Quantum walks from the wave equation} \label{sec:QW}

Now we consider the wave equation instead of the diffusion equation, given by
\[
\ddot{u}
= \f{1}{4d} \Delta u
\]
over $\real^d$.
Equivalently,
\begin{align} \label{eq:wave}
\begin{split}
\dot{u} = v, \qquad
\dot{v} = \f{1}{4d} \Delta u.
\end{split}
\end{align}
We will consider a particular\footnote{There are alternative options, such as $u(n+1) = u(n) + v(n)$, $v(n+1) = v(n) - (I-P^2) u(n)$.} discrete space and time approximation of the diffusion equation, given by
\begin{align} \label{eq:discrete-wave}
\begin{split}
\fo t\in\ZZ_+,\qquad
u(n+1) &= P u(n) + v(n) \\
v(n+1) &= P v(n) - (I - P^2) u(n)
\end{split}
\end{align}
for $u(n),v(n) \in \ell_2(\ZZ^d)$.
The following exposition largely follows \cite{taoblog}.

\paragraph{Discrete approximation.}
We can again motivate this approximation by considering the case with $P$ the random walk transition matrix on $\integer^d$.
We introduce a small parameter $\epsilon>0$ and replace the time set $\ZZ$ by $\epsilon\ZZ$ and the lattice $\ZZ^d$ by $\epsilon \ZZ^d$.
Consider then the evolution
\begin{align} \label{eq:discrete-wave-eps}
\begin{split}
\fo t\in\epsilon\ZZ_+,\qquad
u_\epsilon(n+1) &= P_\epsilon u_\epsilon(n) + \epsilon v_\epsilon(n) \\
v_\epsilon(n+1) &= P_\epsilon v_\epsilon(n) - \frac{1}{\epsilon} (I - P_\epsilon^2) u_\epsilon(n)
\end{split}
\end{align}
where $P_{\epsilon}$ is the random walk transition matrix on $\epsilon \integer^d$.
This evolution coincides with \eqref{eq:discrete-wave} for~$\epsilon=1$.
Now assume e.g.\ that the initial condition $(u(0),v(0))$ of \eqref{eq:wave} is continuous and has a compact support and that the initial condition of \eqref{eq:discrete-wave} is just the restriction of $(u(0),v(0))$ to $\epsilon\ZZ^d$.
Then the following again holds: for any given $(t,x)\in\RR_+\times\RR^d$, we have
\[
\lim_{\epsilon\ri0_+} u_\epsilon(t_\epsilon,x_\epsilon) = u(t,x), \qquad
\lim_{\epsilon\ri0_+} v_\epsilon(t_\epsilon,x_\epsilon) = v(t,x)
\]
where for any $\epsilon>0$, $t_\epsilon\in\epsilon\ZZ_+$ and $x_\epsilon\in \epsilon\ZZ^d$ are such that $\lim_{\epsilon\ri0_+}t_\epsilon = t$ and $\lim_{\epsilon\ri0_+}x_\epsilon = x$.

This argument motivates identifying the dynamics \eqref{eq:discrete-wave} on $\integer^d$ with the discrete wave equation on $\integer^d$.
More generally, we identify \eqref{eq:discrete-wave} with $P$ the random walk transition matrix on a general graph with the discrete wave equation on that graph.

\paragraph*{Unitary dynamics.}
It is easy to check that the following ``energy'' quantity
\[
\| \sqrt{I-P^2} \, u(n) \|^2 + \| v(n) \|^2
\]
is conserved.
Indeed, if we define $w(n) = \sqrt{I-P^2} u(n)$, then we can equivalently write~\eqref{eq:discrete-wave} as
\begin{align*}
w(n+1) &= P w(n) + \sqrt{I-P^2} v(n) \\
v(n+1) &= P v(n) - \sqrt{I - P^2} w(n).
\end{align*}
The corresponding propagator
\begin{equation} \label{eq:unitary}
U
= \begin{bmatrix}
P & \sqrt{I - P^2} \\
- \sqrt{I - P^2} & P
\end{bmatrix}
\end{equation}
is then \emph{unitary}, and hence conserves the norm or energy $\| w(n) \|^2 + \| v(n) \|^2$.
Such a unitary that encodes an operator $P$ as one of its blocks is called a ``unitary dilation'' in the mathematics literature.

The $n$-step propagator $U^n$ can be neatly expressed as
\begin{equation} \label{eq:Un}
U^n
= \begin{bmatrix}
T_n(P) & \sqrt{I-P^2} \, U_{n-1}(P) \\
- \sqrt{I-P^2} \, U_{n-1}(P) & T_n(P)
\end{bmatrix}.
\end{equation}
where $(T_n)_{n\in\ZZ_+}$ and $(U_n)_{n\in\ZZ_+}$ are respectively the families of Chebyshev polynomials of the first and second kind.

\subsection{Quantum walks and block encodings} \label{sec:QWs-block-encoding}

Arguably, such unitary dynamics are a natural map between quantum states, in which case we propose to call the resulting dynamics a \emph{quantum walk} on the underlying graph.
In fact, the unitary dilation in \eqref{eq:unitary} is bread and butter in the quantum computing literature -- there it is usually called a \emph{block encoding} \cite{chakraborty2018power} of the operator $P$.
Such block encodings are central to our current understanding of quantum algorithms -- see e.g.~the recent ``grand unification of quantum algorithms'' in the form of the quantum singular value transformation \cite{gilyen2019quantum}.
One of the first (if not the first) demonstrations of such block encoding was in the seminal work on quantum walks by Szegedy \cite{szegedy2004quantum}.
Szegedy constructed a unitary quantum walk operator $W$, different from the propagator~$U$ in \eqref{eq:unitary}, but resembling it in that
\begin{enumerate}
\item
$W$ is a unitary dilation of the random walk operator $P$, and
\item 
$W^n$ is a unitary dilation of $T_n(P)$.
\end{enumerate}
Szegedy used some additional properties of their particular construction of $W$.
However, we now understand that any operator satisfying the properties 1.~and 2.~can be used as a quantum walk operator, allowing one to reproduce the algorithmic results on quantum walks referenced earlier \cite{apers2019unified}.
This motivates our interpretation of the discrete wave dynamics as a quantum walk -- in fact, it is surprising that this connection seems to not have been made earlier.

We note that, while the $u$- and $v$-components (or $w$- and $v$ components) could be interpreted as position and momentum in the original wave equation, they lose that meaning in the quantum walk interpretation.
Rather, the two components should be interpreted as the components of a state with two chiralities.

\subsection{Varopoulos-Carne bound and quantum fast-forwarding}

A nice demonstration of this connection is the use of a particular identity for Chebyshev polynomials in random walk theory on the one hand (also discussed in \cite{taoblog}), and in quantum algorithms design on the other hand.
The identity is the following: for any self-adjoint contraction $P$ it holds that
\begin{equation} \label{eq:cheb-binom}
P^n
= \sum_{k \in \integer} q_n(k) T_{|k|}(P),
\end{equation}
where $q_n(k)$ is the probability that a random walk on the line, starting from the origin, is at position~$k$ after $n$ steps.

In random walk theory, this identity is used to prove the \emph{Varopoulos-Carne bound} (see \cite{varopoulos1985long,carne1985transmutation} and a nice discussion in \cite{lyons2017probability}).
The bound states that for any pair of vertices $x,y$ it holds that
\[
P^n(x,y)
\leq 2 \exp(-d(x,y)^2/(2n)),
\]
where $d(x,y)$ is the shortest path distance between vertices $x$ and $y$.
The main idea of the bound is to apply Hoeffding's bound to the coefficients $q_n(k)$, showing that
\begin{equation} \label{eq:cheb-approx}
P^n
\approx \sum_{|k| \leq O(\sqrt{n})} q_n(k) T_{|k|}(P).
\end{equation}
Now notice that $[T_{|k|}(P)](x,y) = 0$ if $d(x,y) > k$, and so $P^n(x,y) \approx 0$ when $d(x,y) > O(\sqrt{n})$.

In quantum walk algorithms, the identities \eqref{eq:cheb-binom} and \eqref{eq:cheb-approx} were used recently in a technique called \emph{quantum fast-forwarding} \cite{apers2019quantum}.
Combining the expression \eqref{eq:Un} for the quantum walk propagator with the approximation \eqref{eq:cheb-approx}, we see that
\begin{equation} \label{eq:trunc-cheb}
\sum_{|k| \leq O(\sqrt{n})} q_n(k) U^{|k|}
\approx \begin{bmatrix} P^n & \star \\ \star & P^n \end{bmatrix},
\end{equation}
where the $\star$ represent non-relevant entries.
Through an additional quantum algorithmic technique called ``linear combination of unitaries'' \cite{childs2012hamiltonian}, it is possible to algorithmically implement a unitary dilation of \eqref{eq:trunc-cheb} (and hence of $P^n$) while effectively only implementing $O(\sqrt{n})$ quantum walk steps.
This explains the terminology ``quantum fast-forwarding''.

\section{Quantum walks on lattices}

The preceding discussion shows that we can understand quantum walks through the lense of Chebyshev polynomials.
We demonstrate this connection by studying such Chebyshev polynomials for the particular case of lattices.
Specifically, we are interested in the dynamics
\[
U^n \begin{bmatrix} e_0 \\ 0 \end{bmatrix}
= \begin{bmatrix} T_n(P) \, e_0 \\ - \sqrt{I-P^2} \, U_{n-1}(P) \, e_0 \end{bmatrix},
\]
where $e_0$ is the indicator on some vertex of the lattice.
This corresponds to the discrete wave equation with initial conditions $u(0) = e_0$ and $v(0) = 0$.
Equivalently, it corresponds to an $n$-step quantum walk starting from $e_0$.
We will be interested in the spreading behavior of the quantum walk, and will focus on the $u$-component of the resulting state, i.e., the vector $T_n(P) e_0$.
This vector induces a (subnormalized) measure $\nu$ on the lattice by setting $\nu_x = \left| [T_n(P)]_{0,x} \right|^2$, and this measure will be the object of our study.
Operationally, it corresponds to the probability of measuring the quantum walk in a certain chirality at vertex $x$ (see discussion at the end of \cref{sec:QWs-block-encoding}).

\subsection{Quantum walk on the line} \label{sec:QW-line}

On the infinite line, the $T_n(P)$ component of the quantum walk takes a particularly simple form.
Specifically, consider the line with set of vertices $\integer$ and commuting operators $Q_1$ and $Q_1^{-1}$ defined by $Q_1 e_x = e_{x+1}$ and $Q_1 ^{-1} e_{x} = e_{x-1}$, where for any $x\in\ZZ$, $e_x$ is the function on $\ZZ$ taking the value 1 at $x$ and 0 elsewhere.
The random walk transition matrix is $P = (Q_1 + Q_1^{-1})/2$, and the random walk distribution $[P^n]_{0,\cdot}$ after $n$ steps essentially corresponds to a binomial distribution around~$0$ with standard deviation $O(\sqrt{n})$.
In contrast, if we use the formula $T_n((x+x^{-1})/2) = (x^n + x^{-n})/2$ we can rewrite the Chebyshev polynomial $T_n(P)$ as
\begin{equation} \label{eq:ch-cycle}
T_n(P)
= T_n\left(\frac{Q_1+Q_1^{-1}}{2}\right)
= \frac{Q_1^n + Q_1^{-n}}{2}.
\end{equation}
This implies that $[T_n(P)]_{0,x} = e_n(x)/2 + e_{-n}(x)/2$, and so the quantum walk shows two peaks precisely at a distance $n$ from the origin.
These dynamics represent the \emph{ballistic} nature of the quantum walk dynamics, as opposed to the \emph{diffusive} nature of the random walk dynamics.
The hope is that this improved spreading rate persists in more general graphs as well.

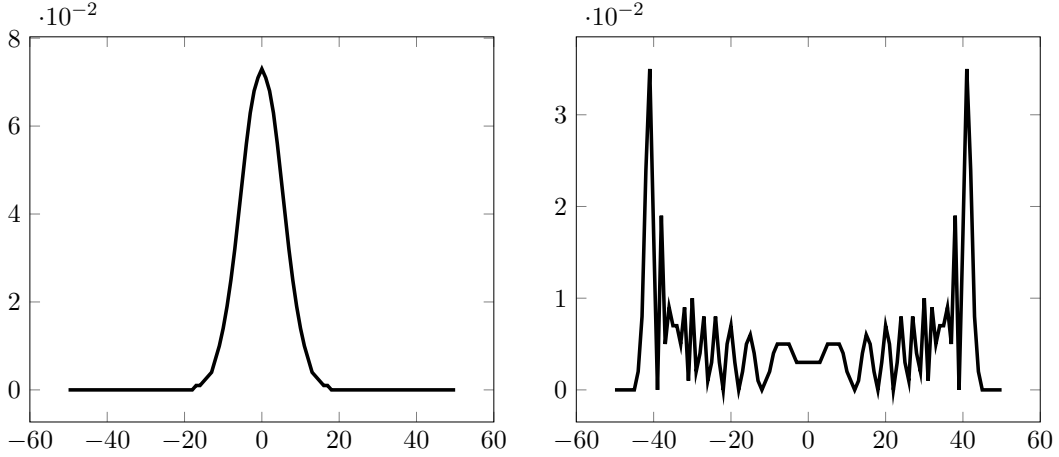
\begin{figure}[htb]
\centering
\subfigure{
    \begin{tikzpicture}[scale=0.9]
    \begin{axis}
    \addplot [black, sharp plot, line width = 1.5pt] table [col sep=comma, x=X, y=Y] {RW-Z.csv};
    \end{axis}
    \end{tikzpicture}
}
\subfigure{
    \begin{tikzpicture}[scale=0.9]
    \begin{axis}
    \addplot [black, sharp plot, line width = 1.5pt] table [col sep=comma, x=X, y=Y] {chebyshev-Z.csv};
    \end{axis}
    \end{tikzpicture}
}
\caption{(left) Probability measure induced by $[P_L^n]_{0,\cdot}$ for $n = 50$, corresponding to an $n$-step lazy random walk on the line. (right) Probability measure induced by $[T_n(P_L)]_{0,\cdot}^2$ for $n = 50$, associated to an $n$-step quantum walk on the line. (left and right) Lines drawn continuously for clarity.} \label{fig:RW-cheb}
\end{figure}

If instead of the simple random walk we consider the \emph{lazy} random walk on the cycle, $P_L = (I + P)/2$, there is no longer a simple expansion of $T_n(P_L)$.
Instead, we retrieve the fickle (but still ballistic) behavior usually associated to the quantum walk on the line.
The behaviors of $P_L^n$ and $T_n(P_L)$ are illustrated in \cref{fig:RW-cheb}.

\subsection{Quantum walk on the lattice}

\begin{figure}[htb]
\centering
\subfigure{
    \begin{tikzpicture}[scale=0.9]
    \begin{axis}[view={0}{90},
                    colormap={whiteblue}{color=(white) color=(black) color=(black)}]
        \addplot3[surf, mesh/rows=71, mesh/cols=71, opacity=0.01,fill opacity=0.5] table {RW-Z2.csv};
    \end{axis}
    \end{tikzpicture}
}
\subfigure{
    \begin{tikzpicture}[scale=.9]
    \begin{axis}[view={0}{90},
                    colorbar,
                    colormap={whiteblue}{color=(white) color=(black) color=(black)}]
        \addplot3[surf, mesh/rows=71, mesh/cols=71, opacity=0.01,fill opacity=0.5] table {cheb-Z2.csv};
    \end{axis}
    \end{tikzpicture}
}
\caption{(left) Probability measure induced by $[P^n]_{0,\cdot}$ for $n = 50$, corresponding to an $n$-step random walk on the lattice. (right) Probability measure induced by $[T_n(P)]_{0,\cdot}^2$ for $n = 50$, associated to an $n$-step quantum walk on the lattice.} \label{fig:RW-cheb-Z2}
\end{figure}
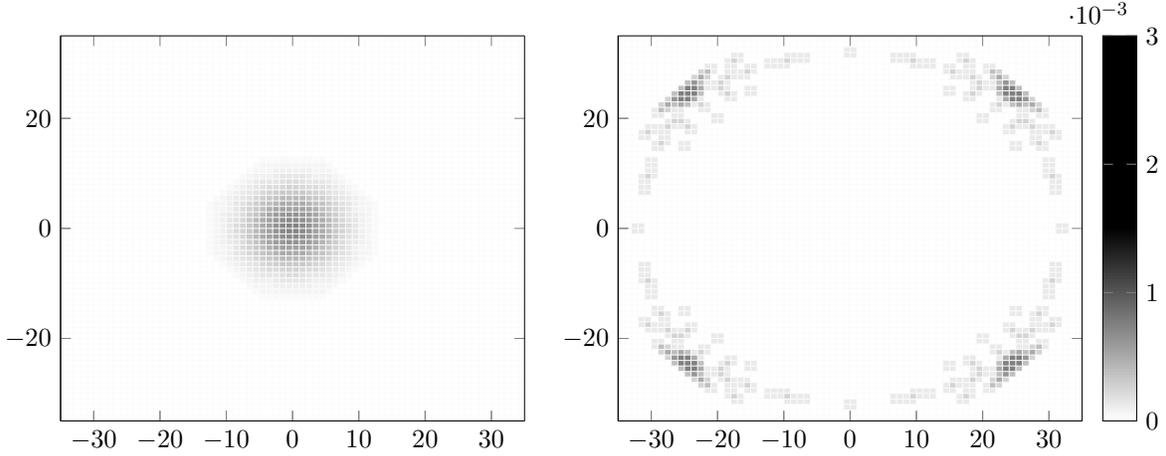

As a more involved example, consider a random walk on the $2$-dimensional lattice with vertices $V = \{(x,y) \in \integer^2\}$ and commuting generators $Q_1,Q_1^{-1},Q_2,Q_2^{-1}$ defined by $Q_1 e_{x,y} = e_{x+1,y}$ and $Q_2 e_{x,y} = e_{x,y+1}$, and similarly for the inverses.
If we set $P_1 = (Q_1+Q^{-1}_1)/2$ and $P_2 = (Q_2+Q^{-1}_2)/2$, then the random walk transition matrix is given by
\[
P
= \frac{P_1 + P_2}{2}.
\]
In \cref{fig:RW-cheb-Z2} we compare the measure induced by the random walk transition matrix $P^n$ and that of the Chebyshev polynomial $T_n(P)$.

To get a handle on the operator $T_n(P)$, we will use the following expansion of a degree-$n$ bivariate Chebyshev polynomial into univariate degree-$n$ Chebyshev polynomials:
\begin{equation} \label{eq:bivar-exp}
T_n\left(\frac{X + Y}{2}\right)
= \sum_{(p,q) \in \lin 0,n\rin} a_{n,p,q} T_p(X) T_q(Y),
\end{equation}
for a set of real coefficients $(a_{n,p,q})_{p,q\in\lin 0,n\rin}$.
Since $T_n(P) = T_n((P_1+P_2)/2)$ is a degree-$n$ polynomial in the commuting operators $P_1$ and $P_2$, we can use the same expansion to rewrite
\begin{align*}
T_n\left(\frac{P_1 + P_2}{2}\right)
&= \sum_{(p,q) \in \lin 0,n\rin} a_{n,p,q} T_p(P_1) T_q(P_2) \\
&= \sum_{(p,q) \in \lin 0,n\rin} a_{n,p,q} \left(\frac{Q^p_1 + Q_1^{-p}}{2}\right)
	\left(\frac{Q^{q}_2 + Q_2^{-q}}{2}\right),
\end{align*}
where we reused the fact (\cref{eq:ch-cycle}) that $T_p(P_1) = (Q_1^p+Q_1^{-p})/2$.
Now notice that
\[
[T_n(P)]_{(0,0),(p,q)}
= a_{n,|p|,|q|}/2^{k(|p|,|q|)},
\]
where $k(|p|,|q|) \in \{0,1,2\}$ denotes the number of nonzero entries among $|p|,|q|$.
We can hence lower bound the probability for a quantum walk, starting from the origin $(0,0)$, to reach a vertex $(p,q)$ by 
\begin{equation} \label{eq:qt-lattice}
\left| [T_n(P)]_{(0,0),(p,q)} \right|^2
= a_{n,|p|,|q|}^2/2^{2k(|p|,|q|)}
\geq a_{n,|p|,|q|}^2/16
\end{equation}
(since $k(p,q)\leq 2$). 
These coefficients are fully determined by the Chebyshev expansion in \eqref{eq:bivar-exp}, and so the problem reduces to bounding these coefficients.
More specifically, we will consider the induced measure $\mu_n$ on $[-1,1]^2$ given by
\bq
\mu_n&\df&\f{ \sum_{p,q} a_{n,|p|,|q|}^2\delta_{(p/n,q/n)}}{ \sum_{p,q} a_{n,|p|,|q|}^2}
\eq
where $\delta_{(p/n,q/n)}$ is the Dirac mass at $(p/n,q/n)\in[0,1]^2$.
As our main technical contribution, we prove a weak limit theorem on $\mu_n$.
\begin{theorem}
There exists a continuous measure $\mu$ on $[-1,1]^2$ such that $\mu_n \rightarrow \mu$.
\end{theorem}

As the most importance consequence of this weak limit, it proves that the quantum walk spreads ballistically on the lattice -- after $n$ steps, the expected distance from the origin is $\Omega(n)$.
Using very different machinery, similar conclusions were found in earlier work \cite{mackay2002quantum,baryshnikov2011two}.

\section{Proof of weak convergence on the lattice}

Our goal is to investigate the weak convergence for $n$ large of $\mu_n$ toward a law $\mu$ on $[0,1]^2$.
To do so, it will be more convenient to consider the measure 
on $[0,1]^2$ given by
\bq
\gamma_n&\df&\sum_{p,q\in\lin 0, n\rin} a_{n,p,q}^2\delta_{(p/n,q/n)}
\eq
\par
We will show there exists a probability measure $\gamma\neq 0$ on $[0,1]^2$ such that  for large $n$, $\gamma_n$ weakly converges toward $\gamma$.
Since for any $n\in\ZZ_+$,
\bq
\mu_n&=&\f{\gamma_n}{\gamma_n([0,1]^2)}\eq
we deduce the weak convergence of $\mu_n$ toward $ \mu\df\gamma/\gamma([0,1]^2)=\gamma$.\par
The method will consist in showing the convergence of the (bivariate) even moments of $\gamma_n$ toward those of $\gamma$.
More precisely, for any $K,L\in\ZZ_+$, consider the polynomial function
\bq
\varphi_{K,L}\st [0,1]^2\ni(x,y)&\mapsto & x^{2K}y^{2L}\eq
We are to find a measure $\gamma\neq 0$ on $[0,1]^2$ such that for any $K,L\in\ZZ_+$,
\bq
\lim_{n\ri\iy} \gamma_n[\varphi_{K,L}]&=&\gamma[\varphi_{K,L}]\eq
\par
Taking into account the Stone-Weierstrass theorem (applied on the compact set $[0,1]^2$ with respect to the algebra generated by the $\varphi_{K,L}$, for $K,L\in\ZZ_+$), this convergence is sufficient to deduce the weak convergence of $\gamma_n$ toward $\gamma$.
\par
The proof is split in three main parts corresponding to the following subsections.

\subsection{Second moment as an integral}

We start by approximating the second moments of $\mu_n$ by some simpler integral.
Consider the function
\begin{equation} \label{eq:def-h}
h(x,y)
\coloneqq T_n\lt(\f{\cos(x)+\cos(y)}2\rt)
= \sum_{p,q\in\lin 0,n\rin} a_{n,p,q} \cos(px)\cos(qy),
\end{equation}
where the last equality follows from \cref{eq:bivar-exp} and the definition of the Chebyshev polynomials.
We show the following approximation.
\begin{proposition} \label{claim:moment-to-integral}
We have
\bq\lim_{n \to \infty} \left|
\gamma_n[\varphi_{K,L}]
- \frac{1}{\pi^2 n^{2(K+L)}} \int_{[0,2\pi]^2}
    \left( \partial^K_x \partial^L_y h(x,y) \right)^2 dxdy
\right|& = &0.
\eq
\end{proposition}
The proof consist of the two following results.
\begin{lemma}\label{lem1}
We have
\bq
\frac{1}{\pi^2 n^{2(K+L)}} \int  \left( \partial^K_x \partial^L_y h(x,y) \right)^2\,dxdy&=&
\frac{1}{ n^{2(K+L)}} \sum_{p,q\in\lin 0,n\rin} a^2_{n,p,q}p^{2K}q^{2L }w_K(p)w_L(q)\\
&=&\gamma_n[\varphi_{K,L}w_K(n\cdot)\otimes w_L(n\cdot)]
\eq
where
\bq
\fo k\in\ZZ_+,\,\fo r\in\ZZ_+,\qquad w_k(r)&\df&\lt\{\begin{array}{ll}
2&\hbox{, if $k$ is even and $r=0$}\\
0&\hbox{, if $k$ is odd and $r=0$}\\
1&\hbox{, otherwise}\end{array}\rt.\eq
\end{lemma}
\proof
We rewrite
\bqn{rhs}
\partial^K_x \partial^L_y h(x,y)
= \sum_{p,q\in\lin 0,n\rin} a_{n,p,q}p^Kq^L \xi_K(px)\xi_L(qy)\eqn
where $\xi_r(z)$ denotes the $r$-th derivative $\cos^{(r)}(z)$.
Now recall the orthogonality relations:
\bq
\fo p,p'\in\ZZ_+,\qquad
\f1{2\pi}\int_0^{2\pi} \xi_r(px)\xi_r(p'x)\, dx
&=& \lt\{
\begin{array}
{ll}
1&\hbox{, if $p=p'=0$ and $r\equiv 0[2]$}\\
1/2&\hbox{, if $p=p'\geq 1$}\\
0&\hbox{, otherwise}
\end{array}\rt.\eq
Taking the square of \eqref{rhs} and integrating, we get
\bq
\lefteqn{\f1{(2\pi)^2}\int \lt(\partial^K_x \partial^L_y h(x,y)\rt)^2dxdy}\\
&=&
\sum_{p,q,p',q'\in\lin 0,n\rin} a_{n,p,q}p^Kq^L a_{n,p',q'}(p')^K(q')^L \f1{(2\pi)^2}\int \xi_K(px)\xi_L(qy)\xi_K(p'x)\xi_L(q'y)\, dxdy\\
&=&\sum_{p,q,p',q'\in\lin 0,n\rin} a_{n,p,q} a_{n,p',q'}p^K(p')^Kq^L(q')^L \lt(\f1{2\pi}\int \xi_K(px)\xi_K(p'x)\, dx\rt)\lt(\f1{2\pi}\int \xi_L(qy)\xi_L(q'y)\, dy\rt)\\
&=&\sum_{p,q\in\lin 0,n\rin} a^2_{n,p,q}p^{2K}q^{2L } \lt(\f1{2\pi}\int \xi_K^2(px)\, dx\rt)\lt(\f1{2\pi}\int \xi_L^2(qy)\, dy\rt)\\
&=&\f14\sum_{p,q\in\lin 0,n\rin} a^2_{n,p,q}p^{2K}q^{2L }w_K(p)w_L(q)
= \f14n^{2(K+L)}\gamma_n[\varphi_{K,L}w_K(n\cdot)\otimes w_L(n\cdot)].
\eq
\wwtbp
\par
We deduce:
\begin{corollary}\label{c1}
We have for any $n\in\ZZ_+$,
\bq
 \sum_{p,q\in\lin 0,n\rin}a^2_{n,p,q}&\leq &4
 \eq
\end{corollary}
\proof
Considering the case $K=L=0$ in Lemma \ref{lem1}, we get
\bq
\f1{\pi^2}\int_{[0,2\pi]^2} T_n^2\lt(\f{\cos(x)+\cos(y)}2\rt)\, dxdy
&=&\sum_{p,q\in\lin 0,n\rin}w_0(p)w_0(q)a_{n,p,q}^2\\
&\geq & \sum_{p,q\in\lin 0,n\rin}a_{n,p,q}^2
\eq
To conclude to the announced bound, note that the l.h.s.\ is bounded above by $4$, since $T_n(z)$ takes its values in $[-1,1]$ for any $z\in[-1,1]$.\wwtbp
\par
The next step enables to end the proof of Proposition \ref{claim:moment-to-integral}.
\begin{lemma} \label{lem:alt-moment}
We have
\bq\lim_{n\ri\iy} \vert \gamma_n[\varphi_{K,L}w_k(n\cdot)\otimes w_L(n\cdot)]-\gamma_n[\varphi_{K,L}]\vert& =& 0\eq
\end{lemma}
\proof
To prove the above convergence, observe that for any given $K,L\in\ZZ_+$,
the quantity
\bq
\vert \gamma_n[\varphi_{K,L}w_k(n\cdot)\otimes w_L(n\cdot)]-\gamma_n[\varphi_{K,L}]\vert&=&\lt\vert
\sum_{p,q\in\ZZ_+}a^2_{n,p,q}\lt(\f{p}{n}\rt)^{2k}\lt(\f{q}{n}\rt)^{2L}(w_K(p)w_L(q)-1)\rt\vert\eq
converges to zero for large $n\in\NN$.
\par\me
Let us consider the term 
\bq
p^{2k}q^{2L}(w_K(p)w_L(q)-1)\eq
It vanishes if either $p\geq 1$ and $q\geq 1$ or $p=0$ and $K\geq 1$ or $q=0$ and $L\geq 1$.
Thus it can be non zero only in the two following (non-disjoint) alternatives:
\par\sm
$\bullet$ If $p=0$ and $K=0$, then its value is
\bq
q^{2L}(2w_L(q)-1)&=&\lt\{\begin{array}{ll}
q^{2L}&\hbox{, if $q\geq 1$}\\
0&\hbox{, if $q=0$ and $L\geq 1$}\\
3&\hbox{, if $q=0$ and $L=0$}\end{array}\rt.\eq
\par\sm
$\bullet$ If $q=0$ and $L=0$, then its value is
\bq
p^{2K}(2w_K(p)-1)&=&\lt\{\begin{array}{ll}
p^{2K}&\hbox{, if $p\geq 1$}\\
0&\hbox{, if $p=0$ and $K\geq 1$}\\
3&\hbox{, if $p=0$ and $K=0$}\end{array}\rt.\eq
\par\sm
It follows that for any $n\in\NN$,
\bq\lefteqn{
\vert \gamma_n[\varphi_{K,L}w_k(n\cdot)\otimes w_L(n\cdot)]-\gamma_n[\varphi_{K,L}]\vert}\\&=&
3a^2_{n,0,0}\un_{\{(0,0)\}}(K,L)+\sum_{q\in\NN} a^2_{n,0,q} \lt(\f{q}{n}\rt)^{2L}\un_{\{0\}}(K)+\sum_{p\in\NN} a^2_{n,p,0} \lt(\f{p}{n}\rt)^{2K}\un_{\{0\}}(L)\\
&=&3a^2_{n,0,0}\un_{\{(0,0)\}}(K,L)+\sum_{q\in\NN} a^2_{n,p,0} \lt(\lt(\f{p}{n}\rt)^{2K}\un_{\{0\}}(L)+\lt(\f{p}{n}\rt)^{2L}\un_{\{0\}}(K)\rt)\\
&\leq &3a^2_{n,0,0}+2\sum_{q\in\NN} a^2_{n,p,0} \\
&\leq & 3\sum_{p\in\ZZ_+} a^2_{n,p,0}
\eq
(where the the symmetry in $(p,q)$ of $a_{n,p,q}$ was taken in account in the second equality).
\par
Thus it is sufficient to show that 
\bqn{but1}
\lim_{n\ri\iy} \sum_{p\in\ZZ_+} a^2_{n,p,0}&=&0\eqn
\par
From the definition of the coefficients $(a_{n,p,q})_{n\in\ZZ_+, p,q\in\lin 0,n\rin}$ we get for any given $n\in \ZZ_+$ and $x\in[0,\pi]$,
\bq
\f1{2\pi}\int_{[0,2\pi]} T_n\lt(\f{\cos(x)+\cos(y)}{2}\rt)\, dy &=&\sum_{p,q\in\lin 0,n\rin} a_{n,p,q}\cos(px)\f1{2\pi}\int_{[0,2\pi]} \cos(qy)\, dy\\
&=&\f12\sum_{p\in\lin 0,n\rin}a_{n,p,0}\cos(px)\eq
\par
Taking the square and integrating, we get
\bq
\f1{2\pi}\int\lt(\f1{2\pi}\int_{[0,2\pi]} T_n\lt(\f{\cos(x)+\cos(y)}{2}\rt)\, dy\rt)^2dx &=&\f{a_{n,0,0}^2}8+\f18\sum_{p\in\lin 0,n\rin} a_{n,p,0}^2\eq
so that \eqref{but1} amounts to 
\bq
\lim_{n\ri\iy} \f1{2\pi}\int_{[0,2\pi]}\lt(\f1{2\pi}\int_{[0,2\pi]} T_n\lt(\f{\cos(x)+\cos(y)}{2}\rt)\, dy\rt)^2dx&=&0
\eq
or, by symmetry, to
\bqn{but2}
\lim_{n\ri\iy}\int_{[0,\pi]}\lt(\int_{[0,\pi]} T_n\lt(\f{\cos(x)+\cos(y)}{2}\rt)\, dy\rt)^2dx&=&0
\eqn
\par
Fix $x\in [0,\pi]$. To investigate the integral $\int_{[0,\pi]} T_n\lt(\f{\cos(x)+\cos(y)}{2}\rt)\, dy$ we consider the
change of variable $y\mapsto \theta_x(y)$ where $\theta_x(y)$ is the unique solution belonging to $[0,\pi]$ of the equation
\bq
\cos(\theta_x(y))&=&\f{\cos(x)+\cos(y)}{2}\eq
\par
To simplify notations, from now on we remove the index $x$.
The mapping $\theta$ is a bijection from $[0,\pi]$ to 
\bq
I&\df& \lt[\arccos\lt(\f{\cos(x)+1}{2}\rt),\arccos\lt(\f{\cos(x)-1}{2}\rt)\rt]\eq
and 
we have \bq
\fo y\in (0,\pi),\qquad \vert \theta'(y)\vert&=&\f12\lt\vert \f{\sin(y)}{\sin(\theta(y))}\rt\vert\eq
\par
The change of variable formula implies that for any measurable function $f\in \LL^1(\ell_I)$ (where $\ell_I$ is the restriction of the Lebesgue measure on $I$),
\bqn{cv1}
\f12\int_{[0,\pi]}f(\theta(y))\vert \theta'(y)\vert\, dy &=& \int_I f(z)\, dz\eqn
\par
Taking into account that for $y\in[0,\pi]$, we have
\bq
\vert \sin(y)\vert &=&\sqrt{1-\cos^2(y)}\\
&=&\sqrt{1-(2\cos(\theta(y))-\cos(x))^2}\eq 
we can write \eqref{cv1} as
\bq
\f12\int_{[0,\pi]}f(\theta(y))\f{\sqrt{1-(2\cos(\theta(y))-\cos(x))^2}}{\vert\sin(\theta(y))\vert}\, dy &=& \int_I f(z)\, dz\eq
\par
Define the function $g$ on $I$
via
\bq
\fo z\in I,\qquad g(z)&\df& f(z) \f{\sqrt{1-(2\cos(z)-\cos(x))^2}}{\vert\sin(z)\vert}\eq
to obtain 
\bq
\f12\int_{[0,\pi]}g(\theta(y))\, dy &=& \int_I g(z)\f{\vert\sin(z)\vert}{\sqrt{1-(2\cos(z)-\cos(x))^2}}\, dz\eq
equality which is valid as soon as $g\circ \theta\in\LL^1(\ell_{[0,\pi]})$.\par
In particular, taking $g=T_n$, which satisfies this condition, we get
\bq
\int_{[0,\pi]}T_n(\theta(y))\, dy &=&2\int_I T_n(z)\f{\vert\sin(z)\vert}{\sqrt{1-(2\cos(z)-\cos(x))^2}}\, dz\eq
namely
\bq
\int_{[0,\pi]}T_n\lt(\f{\cos(x)+\cos(y)}{2}\rt)\, dy &=&2\int_I \cos(nz)\f{\vert\sin(z)\vert}{\sqrt{1-(2\cos(z)-\cos(x))^2}}\, dz\eq
\par
Considering this identity with $n=0$, we get that
the mapping 
\bq
I\ni z&\mapsto &\f{\vert\sin(z)\vert}{\sqrt{1-(2\cos(z)-\cos(x))^2}}\eq
belongs to $\LL^1(\ell_I)$.
Thus we can apply Riemann-Lebesgue theorem to deduce 
\bq
\lim_{n\ri\iy} \int_I \cos(nz)\f{\vert\sin(z)\vert}{\sqrt{1-(2\cos(z)-\cos(x))^2}}\, dz&=&0\eq
and by consequence 
\bq
\lim_{n\ri\iy} \int_{[0,\pi]}T_n\lt(\f{\cos(x)+\cos(y)}{2}\rt)\, dy &=&0\eq
\par
In addition, we have
\bq
\fo n\in\ZZ_+,\fo x\in[0,\pi],\qquad
\lt\vert\int_{[0,\pi]}T_n\lt(\f{\cos(x)+\cos(y)}{2}\rt)\, dy\rt\vert
&\leq & \pi\eq
\par
It follows that \eqref{but2} is a consequence of the dominated convergence theorem.
\wwtbp

\subsection{Negligible boundary region}

For technical reasons that appear later, we wish to limit the integral in \cref{claim:moment-to-integral} to a subdomain $A_\epsilon \subset [0,2\pi]^2$, which for any $\epsilon\in (0,1)$ is defined by
\bq
A_\epsilon&\df& \lt\{(x,y)\in[0,2\pi]^2\st\lve  \f{\cos(x)+\cos(y)}{2}\rve \leq 1-\epsilon \rt\}.
\eq
This is justified since we have
\bq
\lim_{\epsilon\ri 0_+}\lim_{n\ri\iy} \lt\vert \gamma_n[\varphi_{K,L}]-\frac{1}{\pi^2 n^{2(K+L)}} \int_{A_\epsilon}
\left( \partial^K_x \partial^L_y h(x,y) \right)^2 dxdy
\rt\vert&=&0.\eq
This is a consequence of the following lemma, which shows that the contribution of the boundary region $[0,2\pi]^2 \backslash A_\epsilon$ is negligible.
\begin{lemma}
For $\epsilon\in (0,1)$, denote $B_\epsilon\df [0,2\pi]^2\setminus A_\epsilon$.
We have
\bq
\lim_{\epsilon\ri 0_+}\sup_{n\geq  1} \frac{1}{n^{2(K+L)}} \int_{B_\epsilon} \lt(\sum_{p,q\in\lin 0,n\rin} a_{n,p,q}p^Kq^L \xi_K(px)\xi_L(qy)\rt)^2dxdy&=&0
\eq
\end{lemma}
\proof
Note that $B_\epsilon \subset B^-_\epsilon\cup B^+_\epsilon$ with
\bq
B_\epsilon^-&\df& \{(x,y)\in[0,2\pi]^2\st \cos(x)\leq -1+2\epsilon,\, \cos(y)\leq -1+2\epsilon\}\\
B_\epsilon^+&\df& \{(x,y)\in[0,2\pi]^2\st \cos(x)\geq 1-2\epsilon,\, \cos(y)\geq 1-2\epsilon\}\eq
We will prove the statement for the $B_\epsilon^+$ region, since the proof for $B_\epsilon^-$ is analogous.
\par
Define $\eta$ as the unique solution in $(0,\pi)$ of $\cos(\eta)=1-2\epsilon$, we have that $\eta$ goes to $0_+$ when $\epsilon$ goes to $0_+$, and we can rewrite
\bq
\lefteqn{\int_{B_\epsilon^+} \lt(\sum_{p,q\in\lin 0,n\rin} a_{n,p,q}p^Kq^L \xi_K(px)\xi_L(qy)\rt)^2dxdy}\\&=&
\int_{[-\eta,\eta]^2} \lt(\sum_{p,q\in\lin 0,n\rin} a_{n,p,q}p^Kq^L \xi_K(px)\xi_L(qy)\rt)^2dxdy\\
&=&\sum_{p,q,p',q'\in\lin 0,n\rin}a_{n,p,q}a_{n,p',q'}p^Kq^L (p')^K(q')^L \int_{-\eta}^{\eta} 
\xi_K(px)\xi_K(p'x)\, dx \int_{-\eta}^{\eta} 
\xi_L(qy))\xi_L(q'y)\, dy
\eq
\par
\noindent
Let us compute the integrals of the r.h.s.
When $K$ is even, we have
\bq
\int_{-\eta}^{\eta} \xi_K(px)\xi_K(p'x)\, dx
&=&2\int_{0}^{\eta} \cos(px)\cos(p'x)\, dx\\
&=&\int_{0}^{\eta} \cos((p+p')x)+\cos((p-p')x)\, dx\\
&=&-\f{\sin((p+p')\eta)}{p+p'}-\f{\sin((p-p')\eta)}{p-p'}
\eq
\noindent
with the usual convention that $\f{\sin((p-p')\eta)}{p-p'}=\eta$ when $p-p'=0$.
When $K$ is odd, a similar computation shows that
\bq
\int_{-\eta}^{\eta} \xi_K(px)\xi_K(p'x)\, dx&=&\f{\sin((p+p')\eta)}{p+p'}-\f{\sin((p-p')\eta)}{p-p'}.
\eq
\par\sm
\noindent
It follows that the main integral is a sum of terms of the form
\bq
\sum_{p,q,p',q'\in\lin 0,n\rin}\pm a_{n,p,q}a_{n,p',q'}p^Kq^L (p')^K(q')^L \f{\sin((p\pm p')\eta)}{p\pm p'}\f{\sin((q\pm q')\eta)}{q\pm q'},
\eq
for different choices of the signs.
We consider the case with all minus signs, as the other cases follow by essentially the same argument.
By a triangle inequality and the Cauchy-Schwarz inequality we can bound the sum by
\bq
\sum_{p,q,p',q'}a_{n,p,q}&a_{n,p',q'}&\lt\vert \f{\sin((p-p')\eta)}{p-p'}\f{\sin((q-q')\eta)}{q-q'}\rt\vert \\
&\leq &\sqrt{ \sum_{p,q,p',q'}a^2_{n,p,q}\lt(\f{\sin((p-p')\eta)}{p-p'}\f{\sin((q-q')\eta)}{q-q'}\rt)^2}\\&&\quad \sqrt{ \sum_{p,q,p',q'}a^2_{n,p',q'}\lt(\f{\sin((p-p')\eta)}{p-p'}\f{\sin((q-q')\eta)}{q-q'}\rt)^2}\\
&=&\sum_{p,q,p',q'}a^2_{n,p,q}\lt(\f{\sin((p-p')\eta)}{p-p'}\f{\sin((q-q')\eta)}{q-q'}\rt)^2\\
&\leq & 
\sum_{p,q}a^2_{n,p,q} \sum_{r,s\in\ZZ}\lt(\f{\sin(r\eta)}{r}\f{\sin(s\eta)}{s}\rt)^2\\
&=&\sum_{p,q}a^2_{n,p,q} \lt(\sum_{r\in\ZZ}\lt(\f{\sin(r\eta)}{r}\rt)^2\rt)^2\\
&=&\sum_{p,q}a^2_{n,p,q} \lt(\eta^2+2\sum_{r\in\NN}\lt(\f{\sin(r\eta)}{r}\rt)^2\rt)^2\\
&\leq & 4\lt(\eta^2+2\sum_{r\in\NN}\lt(\f{\sin(r\eta)}{r}\rt)^2\rt)^2
\eq
where we used Corollary \ref{c1}.\par
Dominated convergence now implies that
\bq
\lim_{\eta\ri 0_+}\sum_{r\in\NN}\lt(\f{\sin(r\eta)}{r}\rt)^2&=&0,\eq
which proves the claimed convergence.
\wwtbp

\subsection{Simplifying the integral}

We will now approximate the integral $\int_{A_\epsilon} \lt(\partial^K_x \partial^L_y h(x,y)\rt)^2dxdy$ by a (arguably) simpler integral.

\subsubsection{First simplification}
We use Faà di Bruno's formula to rewrite $\partial_x^K h(x,y) = \partial_x^K T_n((\cos(x)+\cos(y))/2)$ as
\bq
\sum_{m_1+2m_2+\cdots +Km_K=K}\f{K!}{m_1!m_2!\cdots m_K!}T_n^{(\sum m_\ell)}
\lt(\f{\cos(x)+\cos(y)}2\rt)\prod_{k\in\lin K\rin}\lt(\f{\xi_{k}(x)}{2k!}\rt)^{m_k},\eq
where the sum is over all nonnegative integers $(m_1, ..., m_K)$ satisfying the constraint $m_1+2m_2+\cdots +Km_K=K$.
Applying Faà di Bruno's formula again we can rewrite $\partial_x^K \partial_y^L h(x,y)$ as
\bqn{Hn}
\nonumber&&\sum_{\begin{array}{c}\scriptscriptstyle m_1+2m_2+\cdots +Km_K=K\\[-2mm]
\scriptscriptstyle n_1+2n_2+\cdots +Ln_L=L
\end{array}}
\f{K!}{m_1!m_2!\cdots m_K!}\f{L!}{n_1!n_2!\cdots n_L!}\\&&T_n^{(\sum m_\ell + \sum n_\ell)}
\lt(\f{\cos(x)+\cos(y)}2\rt)\prod_{k\in\lin K\rin}\lt(\f{\xi_{k}(x)}{2k!}\rt)^{m_k}\prod_{l\in\lin L\rin}\lt(\f{\xi_{l}(y)}{2l!}\rt)^{n_l}\eqn
\par
\noindent
How terrible this expression may seem, as $n \to \infty$ the final integral will be dominated by a single term in the summand.
To see this, we use the following bound.
\begin{lemma}\label{eps}
For any $\epsilon\in (0,1)$ and $r\in\NN$, there exists a constant $ C(\epsilon,r)>0$ such that
\bq
\fo n\in\ZZ_+,\,\fo x\in[-1+\epsilon,1-\epsilon],\qquad \vert T_n^{(r)}(x)\vert&\leq & C(\epsilon,r)n^r\eq
\end{lemma}
\proof
Recall that $T_n(x) = \cos(n\arccos(x))$ for all $x \in [-1,1]$.
The desired result follows from the fact that  any given  derivative of $\arccos$ is bounded on $[-1+\epsilon, 1-\epsilon]$.\wwtbp
\par
It follows that generically in \eqref{Hn}, as $n \rightarrow \infty$, the main term corresponds to $m_1=K$, $n_1=L$, and all other coefficients equal to zero.
We can prove the following lemma, where we define the integral
\bq
I_{K,L}(\epsilon,n)&\df&
\f1{\pi^2n^{2(K+L)}}\int_{A_\epsilon}  \lt(T_n^{(K+L)}\lt(\f{\cos(x)+\cos(y)}{2}\rt)\lt(\f{-\sin(x)}{2}\rt)^K\lt(\f{-\sin(y)}{2}\rt)^L\rt)^2\, dxdy.\eq
\par

\begin{lemma} \label{lem:limit-bruno}
For any $\epsilon>0$ we have 
\bq
\lim_{n\ri\iy} \f1{\pi^2n^{2(K+L)}}\int_{A_\epsilon} \left( \partial_x^K \partial_y^L h(x,y) \right)^2\, dxdy
- I_{K,L}(\epsilon,n)
&=&0.
\eq
\end{lemma}
\proof
Using the expression \eqref{Hn} and expanding the square, up to some factor not depending on $n$, we end up with products of the form
\bqn{TT3}
T_n^{(\sum m_\ell + \sum n_\ell)}
\lt(\f{\cos(x)+\cos(y)}2\rt)T_n^{(\sum m'_\ell + \sum n'_\ell)}
\lt(\f{\cos(x)+\cos(y)}2\rt)\eqn
where $\sum_{\ell=1}^K \ell m_\ell = \sum_{\ell=1}^K \ell m'_\ell = K$ and $\sum_{\ell=1}^L \ell n_\ell = \sum_{\ell=1}^L\ell n'_\ell = L$.
According to Lemma \ref{eps}, up to a constant depending on $\epsilon$, $K$ and $L$, the quantity \eqref{TT3} is bounded independently for $(x,y)\in A_\epsilon$ by 
\bq
n^{\sum m_\ell + m'_\ell + \sum n_\ell + n'_\ell}
\eq
which is negligible with respect to $n^{2(K+L)}$ unless the only nonzero coefficients are $m_1=m'_1=K$ and $n_1=n'_1=L$.
The only potentially non-vanishing term in the integral is hence
\bq
\left( T_n^{(K+L)}\lt(\f{\cos(x)+\cos(y)}{2}\rt)\lt(\f{-\sin(x)}{2}\rt)^K\lt(\f{-\sin(y)}{2}\rt)^L \right)^2
\eq
and this leads to the announced convergence.
\wwtbp
\par

\subsubsection{Second simplification}

We do a second simplification by approximating the integral $I_{K,L}(\epsilon,n)$.
Applying Faà di Bruno's formula to the composition of $T_n$ with cosine, 
we get for any $\theta\in[0,\pi]$,
\bq
\pa^{K+L}_\theta T_{n}(\cos(\theta))=
\!\!\!\!\!\!\!\!\!\!\!\sum_{m_1+2m_2+\cdots +(K+L)m_{K+L}=K+L}\!\! \f{(K+L)!}{m_1!m_2!\cdots m_{K+L}!}T_n^{(m_1+m_2+\cdots+ m_{K+L})}
(\cos(\theta))\!\!\!\!\!\prod_{k\in\lin K+L\rin}\!\!\!\lt(\f{\xi_{k}(\theta)}{k!}\rt)^{\! m_k}\eq
By a similar argument as in \cref{lem:limit-bruno}, we have that when $\theta$ in $(0,\pi)$ and $K,L\in\ZZ_+$ are fixed then
\bqn{beh}
\lim_{n\ri\iy}\lt\vert n^{-(K+L)}\pa^{K+L}_\theta T_n(\cos(\theta))- T_n^{(K+L)}\lt(\cos(\theta)\rt) (-\sin(\theta))^{K+L}\rt\vert&=&0\eqn
Alternatively, taking into account that $T_n(\cos(\theta))=\cos(n\theta)$, we have that
\bq
n^{-(K+L)}\pa^{K+L}_\theta T_n(\cos(\theta))
=n^{-(K+l)}\pa^{K+L}_\theta \cos(n\theta)
= \xi_{K+L}(n\theta)
\eq

\par
Using some uniformity in the behavior \eqref{beh} with respect to $\theta$ away from 0 and $\pi$, as in the proof of \cref{lem:limit-bruno}, we deduce
that for any fixed $\epsilon\in(0,1)$,
\bq
\lim_{n\ri\iy}
\int_{A_\epsilon}\lt\vert \lt(\f{\xi_{K+L}(n\theta(x,y))}{(-\sin(\theta(x,y)))^{K+L}}\rt)^2-n^{-2(K+L)}\lt(T_n^{(K+L)}\lt(\cos(\theta(x,y))\rt)\rt)^2\rt\vert\sin^{2K}(x)\sin^{2L}(y)\, dxdy&=&0
\eq
with $\theta(x,y)$ the unique solution in $[0,\pi]$ of
\bqn{theta}
\cos(\theta(x,y))&=& \f{\cos(x)+\cos(y)}{2}.\eqn
\par
In particular
we have
\bq
\lim_{\epsilon\ri 0_+}\limsup_{n\ri\iy}\vert I_{K,L}(\epsilon,n)- J_{K,L}(\epsilon,n)\vert&=&0\eq
with 
\bq
J_{K,L}(\epsilon,n)&\df&
\f1{\pi^22^{2(K+L)}}\int_{A_\epsilon}  \xi_{K+L}^2(n\theta(x,y))\lt(\f{\sin(x)}{\sin(\theta(x,y))}\rt)^{2K}\lt(\f{\sin(y)}{\sin(\theta(x,y))}\rt)^{2L}\, dxdy\\
&=&\f4{\pi^22^{2(K+L)}}\int_{A_\epsilon\cap[0,\pi]^2}  \xi_{K+L}^2(n\theta(x,y))\lt(\f{\sin(x)}{\sin(\theta(x,y))}\rt)^{2K}\lt(\f{\sin(y)}{\sin(\theta(x,y))}\rt)^{2L}\, dxdy\eq
where symmetries $x\mapsto -x$ and $y\mapsto -y$ were taken into account.

Combining all of the above claims, we get that
\bq
\lim_{\epsilon\ri 0_+}\lim_{n\ri\iy} \lt\vert \gamma_n[\varphi_{K,L}]-J_{K,L}(\epsilon,n)
\rt\vert&=&0.\eq

\subsection{Integral as a second moment}

In the final part of the proof we show that, as $n \to \infty$, the quantity $J_{K,L}(\epsilon,n)$ approximates the second moment of some continuous measure $\gamma$.
More specifically, we prove the following lemma.

\begin{lemma} \label{lem:weak-lim}
It holds that
\[
\lim_{\epsilon\ri 0_+}\limsup_{n\ri\iy} J_{K,L}(\epsilon,n)
= \f1{\pi^2}\int_{[0,\pi]^2}  \lt(\f{\sin(x)}{\sin(\theta(x,y))}\rt)^{2K}\lt(\f{\sin(y)}{\sin(\theta(x,y))}\rt)^{2L}\, dxdy.
\]
\end{lemma}

It follows that we can define the measure $\gamma$ as the image of the measure $\f1{\pi^2} dxdy$ on $[0,\pi]^2$ by the mapping $\Psi$ defined by
\bqn{Psi}
\Psi\,:\,[0,\pi]^2\ni(x,y)&\mapsto& \lt(\f{\sin(x)}{\sin(\theta(x,y))},\f{\sin(y)}{\sin(\theta(x,y))}\rt).\eqn
\par
In particular it appears that $\gamma$ is a probability measure, so that $\mu=\gamma$, as announced previously.\par
Combining the previous arguments, we see that $\lim_{n\to\infty} \gamma_n[\phi_{K,L}] = \gamma[\phi_{K,L}]$, as wanted.
\par
\par\me\noindent\textbf{Proof of Lemma \ref{lem:weak-lim}}\par\sm\noindent
Let us rewrite the integral $J_{K,L}(\epsilon,n)$ using the change of variables
\[
F
\st [0,\pi]^2\ni (x,y)\,\mapsto\, (x,\theta(x,y)).
\]
This is a bijection on its image, which is the set
\bqn{Om}
\Omega
&\df &\{(x,\theta)\in [0,\pi]^2\st 2\cos(\theta)-\cos(x)\in[-1,1]\}.
\eqn
From the change of variables formula (see Lemma \ref{chva2} at the end of this proof), we have that for any measurable $f\st \Omega\ri \RR_+$,
\bqn{chva}\int_{[0,\pi]^2} f(F(x,y)) \, dxdy &=&\int_{\Omega} f(x,\theta)\f{2 \sin(\theta)}{\sqrt{1-(2\cos(\theta)-\cos(x))^2}}\, dxd\theta.\eqn
We can write the integral $J_{K,L}(\epsilon,n)$ in this form by noting that
\bq
\lefteqn{\pi^22^{2(K+L)}J_{K,L}(\epsilon,n)}\\
&=& \int_{\vert \cos(\theta(x,y))\vert \leq 1-\epsilon} \xi_{K+L}^2(n\theta(x,y))\lt(\f{\sin(x)}{\sin(\theta(x,y))}\rt)^{2K}\f{\lt(1-\cos^{2}(y)\rt)^L}{\sin^{2L}(\theta(x,y))}\, dxdy\\
&=&\int_{\vert \cos(\theta(x,y))\vert \leq 1-\epsilon} \xi_{K+L}^2(n\theta(x,y))\lt(\f{\sin(x)}{\sin(\theta(x,y))}\rt)^{2K}\f{\lt(1-(2\cos(\theta(x,y))-\cos(x))^2\rt)^L}{\sin^{2L}(\theta(x,y))}\, dxdy\\
&=&\int_{[0,\pi]^2}f_n(F(x,y))\, dxdy
\eq
where
\bq
\fo (x,\theta)\in\Omega, \qquad f_n(x,\theta)&\df&\un_{[0,1-\epsilon]}(\vert\cos(\theta)\vert)\xi_{K+L}^2(n\theta)\lt(\f{\sin(x)}{\sin(\theta)}\rt)^{2K}\f{\lt(1-(2\cos(\theta)-\cos(x))^2\rt)^L}{\sin^{2L}(\theta)}\eq
\par
The change of variables \eqref{chva}  implies 
\bq
\pi^22^{2(K+L)}J_{K,L}(\epsilon,n)&=&\int_{\Omega} f_n(x,\theta)\f{2 \sin(\theta)}{\sqrt{1-(2\cos(\theta)-\cos(x))^2}}\, dxd\theta\eq\par
\par
Note that
\bq
\fo r\in\ZZ_+,\,\fo z\in[0,\pi],\qquad
 \xi^2_r(z)&\df&
\lt\{\begin{array}{ll}
\f{1+\cos(2z)}2&\hbox{, if $r\equiv 0[2]$}\\[2mm]
\f{1-\cos(2z)}2&\hbox{, if $r\equiv 1[2]$}
\end{array}\rt.\eq
\par
Consider the function $g$ defined on $\Omega$ via
\bq
\fo (x,\theta)\in\Omega, \qquad g(x,\theta)&\df&
\lt(\f{\sin(x)}{\sin(\theta)}\rt)^{2K}\f{\lt(1-(2\cos(\theta)-\cos(x))^2\rt)^L}{\sin^{2L}(\theta)}\f{2 \sin(\theta)}{\sqrt{1-(2\cos(\theta)-\cos(x))^2}}\eq
so that
\bq
2\pi^22^{2(K+L)}J_{K,L}(\epsilon,n)&=&\int_{\Omega} \lt(1+(-1)^{K+L}\cos(2n\theta)\rt)g(x,\theta)\un_{[0,1-\epsilon]}(\cos(\theta))\, dxd\theta\eq\par
For $\epsilon\in(0,1)$ still fixed, the mapping $\Omega\ni(x,\theta)\mapsto g(x,\theta)\un_{[0,1-\epsilon]}(\cos(\theta))$ is integrable with respect to the Lebesgue measure on $\Omega$. We can apply the Riemann-Lebesgue theorem to get
\bq
2\pi^22^{2(K+L)}\lim_{n\ri\iy}J_{K,L}(\epsilon,n)&=&\int_{\Omega}g(x,\theta)\un_{[0,1-\epsilon]}(\cos(\theta))\, dxd\theta,\eq
and by monotone convergence we have
 \bq
 \lim_{\epsilon\ri 0_+} 
\int_{\Omega}g(x,\theta)\un_{[0,1-\epsilon]}(\cos(\theta))\, dxd\theta&=&\int_{\Omega}g(x,\theta)\, dxd\theta.\eq
\par
Finally, note that the previous change of variable computations can be reversed, and yields
\bq
\f1{2\pi^22^{2(K+L)}}\int_{\Omega}g(x,\theta)\, dxd\theta&=&\f1{\pi^2}\int_{[0,\pi]^2}  \lt(\f{\sin(x)}{\sin(\theta(x,y))}\rt)^{2K}\lt(\f{\sin(y)}{\sin(\theta(x,y))}\rt)^{2L}\, dxdy.\eq
This proves  \cref{lem:weak-lim}.
\wwtbp\par
The following result gives the details for the change of variables formula used in Formula \eqref{chva}.
\begin{lemma}\label{chva2}
We have for any measurable $f\st \Omega\ri \RR_+$,
\bq\int_{[0,\pi]^2} f(F(x,y)) \, dxdy &=&\int_{\Omega} f(x,\theta)\f{2 \sin(\theta)}{\sqrt{1-(2\cos(\theta)-\cos(x))^2}}\, dxd\theta.\eq
\end{lemma}
\proof
In addition to the set $\Omega$ defined in \eqref{Om}, for given $\theta\in[0,\pi]$, denote
\bq
\Omega_\theta&\df&\{x\in[0,\pi]\st (x,\theta)\in\Omega\}\eq
\par
Thus for any measurable and non-negative mapping $g\st \Omega\ri\RR_+$, we have by the change of variables formula
\bqn{covf}
\int_{[0,\pi]^2} g(F(x,y)) \vert \dete(\Jac( F))\vert(x,y)\, dxdy &=&\int_{\Omega} g(x,\theta)\, dxd\theta\eqn
where $\Jac( F)$ is the Jacobian matrix associated to $F$.
\par
We have
\bq
\Jac( F(x,y))&=&\lt(\begin{array}{cc}
1&\pa_x\theta(x,y)\\
0&\pa_y\theta(x,y)
\end{array}\rt)
\eq
\par
Differentiating \eqref{theta} with respect to $y$, we get, when $\theta(x,y)\not\in\{0,\pi\}$ (i.e. $(x,y)\not\in \{(0,0), (\pi,\pi)\}$),
\bq
\pa_y\theta(x,y)&=&\f{\sin(y)}{2\sin(\theta(x,y))}\eq
so that
\bq
\vert \dete(\Jac( F))\vert(x,y)&=&\vert \pa_y\theta(x,y)\vert\\
&=&
\f{\vert \sin(y)\vert }{2\vert \sin(\theta(x,y))\vert }\\
&=&\f{\sqrt{1-\cos^2(y)}}{2 \sin(\theta(x,y)) }\\
&=&\f{\sqrt{1-(2\cos(\theta(x,y))-\cos(x))^2}}{2\sin(\theta(x,y)) }\eq
\par
Formula \eqref{covf} rewrites
\bq\int_{[0,\pi]^2} g(F(x,y))\f{\sqrt{1-(2\cos(\theta(x,y))-\cos(x))^2}}{2 \sin(\theta(x,y))} \, dxdy &=&\int_{\Omega} g(x,\theta)\, dxd\theta\eq
or
\bq\int_{[0,\pi]^2} f(F(x,y)) \, dxdy &=&\int_{\Omega} f(x,\theta)\f{2 \sin(\theta)}{\sqrt{1-(2\cos(\theta)-\cos(x))^2}}\, dxd\theta\eq
where $g$ was replaced by the non-negative and measurable mapping $f$ given by
\bq\fo (x,\theta)\in\Omega\setminus \{(0,0), (\pi,\pi)\},\qquad f(x,\theta)&\df& g(x,\theta)\f{\sqrt{1-(2\cos(\theta)-\cos(x))^2}}{2  \sin(\theta) }\eq
\wwtbp
\par
\begin{remark} 
The fact that $\gamma$ is a probability implies (in fact, is equivalent to) the convergence
\bq
\lim_{n\ri\iy}\sum_{p,q\in\lin 0,n\rin} a_{n,p,q}^2&=&\gamma(\un)\ =\ 1
\eq
(compare with Corollary \ref{c1}), which also amounts to
\bq
 \lim_{n\ri\iy}\sum_{p,q\in\lin 1,n\rin} a_{n,p,q}^2&=&1
\eq
according to \eqref{but1}.
\end{remark}\par

\section{Discussion and future work}

We discussed how quantum walks on graphs can be interpreted as solutions of a discrete wave equation on the corresponding graph.
This naturally gives rise to a description of the quantum walk dynamics using Chebyshev polynomials.
We then used the analysis of Chebyshev polynomials to prove a weak limit on the spreading behavior of quantum walks on lattices.

The main open direction of this work is to use Chebyshev polynomials for studying quantum walk spreading and mixing on more general graphs.
The analysis that we did for the lattice seems to naturally extend to lattices of higher dimensions, or Cayley graphs of Abelian groups more generally.
An interesting example out of this category, while still bearing some similarity with the lattice, is the Cayley graph of the Heisenberg group (as studied e.g.~in \cite{bump2017exercise}).
Preliminary simulations indicate a faster spreading rate of quantum walks, yet we are not aware of a proof of this.

Exploring the connection with the wave equation on graphs by Friedman and Tillich~\cite{friedman2004wave} is another open direction.
These authors associate to a Markov  generator $\tr_G$ (so that $I-\tr_G$ is a transition matrix, where $I$ is the identity matrix) on a discrete graph $G$ a diffusion generator $\tr_{\cG}$ on the geometric realization $\cG$ of $G$, where the edges have been replaced by segments of length 1 on which usual one-dimensional Laplacian operators are acting.
The domain of $\tr_{\cG}$ consists of functions satisfying some conditions at the vertices.
Friedman and Tillich then relate the eigenvalues of $\tr_G$ and $\tr_{\cG}$, a feature which enables them to interpret $T_k(I-\tr_G)$ as an operator related to the evolution of the wave equation associated to $\tr_\cG$ (see their Theorem 4.9).
We leave it as an open question whether this interpretation of the action of the Chebyshev polynomials on a transition matrix can be used to deduce our weak convergence result.

\section{Acknowledgements}
Simon Apers is partially supported by French projects EPIQ (ANR-22-PETQ-0007), QUDATA (ANR-18-CE47-0010) and QUOPS (ANR22-CE47-0003-01), and EU project QOPT (QuantERA ERA-NET Cofund 2022-25).

\bibliographystyle{alpha2}
\bibliography{./biblio}

\newcommand{\etalchar}[1]{$^{#1}$}
\newcommand{\arxiv}[1]{arXiv: \href{https://arxiv.org/abs/#1}{\ttfamily{#1}}}
\begin{thebibliography}{AAKV01}

\bibitem[AAKV01]{aharonov2001quantum}
Dorit Aharonov, Andris Ambainis, Julia Kempe, and Umesh Vazirani.
\newblock Quantum walks on graphs.
\newblock In {\em Proceedings of the 33rd {ACM} Symposium on Theory of
  Computing ({STOC})}, pages 50--59. ACM, 2001.
\newblock \arxiv{quant-ph/0012090}.

\bibitem[AGJ21]{apers2019unified}
Simon Apers, Andr{\'a}s Gily{\'e}n, and Stacey Jeffery.
\newblock A unified framework of quantum walk search.
\newblock In {\em Proceedings of the 38th Symposium on Theoretical Aspects of
  Computer Science ({STACS})}. LIPIcs, 2021.
\newblock \arxiv{1912.04233}.

\bibitem[AGJK20]{ambainis2019quadratic}
Andris Ambainis, Andr\'as Gily\'en, Stacey Jeffery, and Martins Kokainis.
\newblock Quadratic speedup for finding marked vertices by quantum walks.
\newblock In {\em Proceedings of the 52nd {ACM} Symposium on Theory of
  Computing ({STOC})}, pages 412--424. ACM, 2020.
\newblock \arxiv{1903.07493}.

\bibitem[Amb03]{ambainis2003quantum}
Andris Ambainis.
\newblock Quantum walks and their algorithmic applications.
\newblock {\em International Journal of Quantum Information}, 1(04):507--518,
  2003.

\bibitem[{A}19]{apers2019qsampling}
Simon Apers.
\newblock Quantum walk sampling by growing seed sets.
\newblock In {\em Proceedings of the 27th European Symposium on Algorithms
  ({ESA})}, volume 144, pages 9:1--9:12. Springer, 2019.
\newblock \arxiv{1904.11446}.

\bibitem[AS19]{apers2019quantum}
Simon Apers and Alain Sarlette.
\newblock Quantum fast-forwarding {Markov} chains and property testing.
\newblock {\em Quantum Information \& Computation}, 19(3\&4):181--213, 2019.
\newblock \arxiv{1804.02321}.

\bibitem[BBBP11]{baryshnikov2011two}
Yuliy Baryshnikov, Wil Brady, Andrew Bressler, and Robin Pemantle.
\newblock Two-dimensional quantum random walk.
\newblock {\em Journal of Statistical Physics}, 142(1):78--107, 2011.

\bibitem[BDH{\etalchar{+}}17]{bump2017exercise}
Daniel Bump, Persi Diaconis, Angela Hicks, Laurent Miclo, and Harold Widom.
\newblock {An exercise (?) in Fourier analysis on the Heisenberg group}.
\newblock In {\em Annales de la Facult{\'e} des sciences de Toulouse:
  Math{\'e}matiques}, volume 26:2, pages 263--288, 2017.

\bibitem[Car85]{carne1985transmutation}
Thomas~Keith Carne.
\newblock {A transmutation formula for Markov chains}.
\newblock {\em Bulletin des sciences math{\'e}matiques (Paris 1985)},
  109(4):399--405, 1985.

\bibitem[CGJ19]{chakraborty2018power}
Shantanav Chakraborty, Andr{\'a}s Gily{\'e}n, and Stacey Jeffery.
\newblock The power of block-encoded matrix powers: improved regression
  techniques via faster {H}amiltonian simulation.
\newblock In {\em Proceedings of the 46th International Colloquium on Automata,
  Languages, and Programming (ICALP)}, pages 33:1--33:14. Schloss
  Dagstuhl--Leibniz-Zentrum fuer Informatik, 2019.
\newblock \arxiv{1804.01973}.

\bibitem[Chi17]{childs2017lecture}
Andrew~M. Childs.
\newblock Lecture notes on quantum algorithms.
\newblock 2017.

\bibitem[CKS17]{childs2017quantum}
Andrew~M. Childs, Robin Kothari, and Rolando~D. Somma.
\newblock Quantum algorithm for systems of linear equations with exponentially
  improved dependence on precision.
\newblock {\em SIAM Journal on Computing}, 46(6):1920--1950, 2017.

\bibitem[CW12]{childs2012hamiltonian}
Andrew~M. Childs and Nathan Wiebe.
\newblock {H}amiltonian simulation using linear combinations of unitary
  operations.
\newblock {\em Quantum Information and Computation}, 12(11\&12):901--924, 2012.

\bibitem[Don52]{donsker1952justification}
Monroe~D. Donsker.
\newblock {Justification and extension of Doob's heuristic approach to the
  Kolmogorov-Smirnov theorems}.
\newblock {\em The Annals of mathematical statistics}, pages 277--281, 1952.

\bibitem[FT04]{friedman2004wave}
Joel Friedman and Jean-Pierre Tillich.
\newblock {Wave equations for graphs and the edge-based Laplacian}.
\newblock {\em Pacific Journal of Mathematics}, 216(2):229--266, 2004.

\bibitem[Gro96]{grover1996fast}
Lov~K. Grover.
\newblock A fast quantum mechanical algorithm for database search.
\newblock In {\em Proceedings of the 28th {ACM} Symposium on Theory of
  Computing ({STOC})}, pages 212--219. ACM, 1996.
\newblock \arxiv{quant-ph/9605043}.

\bibitem[GSLW19]{gilyen2019quantum}
Andr{\'a}s Gily{\'e}n, Yuan Su, Guang~Hao Low, and Nathan Wiebe.
\newblock Quantum singular value transformation and beyond: exponential
  improvements for quantum matrix arithmetics.
\newblock In {\em Proceedings of the 51st {ACM} Symposium on Theory of
  Computing ({STOC})}, pages 193--204. ACM, 2019.
\newblock \arxiv{1806.01838}.

\bibitem[HW20]{harrow2020adaptive}
Aram~W. Harrow and Annie~Y. Wei.
\newblock Adaptive quantum simulated annealing for {B}ayesian inference and
  estimating partition functions.
\newblock In {\em Proceedings of the 14th {ACM-SIAM} Symposium on Discrete
  Algorithms ({SODA})}, pages 193--212. SIAM, 2020.
\newblock \arxiv{1907.09965}.

\bibitem[LP17]{lyons2017probability}
Russell Lyons and Yuval Peres.
\newblock {\em Probability on trees and networks}, volume~42.
\newblock Cambridge University Press, 2017.
\newblock
  {\small\href{http://mypage.iu.edu/~rdlyons/prbtree/book.pdf}{\ttfamily{link}}}.

\bibitem[LPW17]{levin2017markov}
David~A. Levin, Yuval Peres, and Elizabeth~L. Wilmer.
\newblock {\em {M}arkov chains and mixing times}.
\newblock American Mathematical Society, 2017.

\bibitem[MBSS02]{mackay2002quantum}
Troy~D. Mackay, Stephen~D. Bartlett, Leigh~T. Stephenson, and Barry~C. Sanders.
\newblock Quantum walks in higher dimensions.
\newblock {\em Journal of Physics A: Mathematical and General}, 35(12):2745,
  2002.

\bibitem[MP10]{morters2010brownian}
Peter M{\"o}rters and Yuval Peres.
\newblock {\em Brownian motion}, volume~30.
\newblock Cambridge University Press, 2010.

\bibitem[Pid19]{piddock2019quantum}
Stephen Piddock.
\newblock Quantum walk search algorithms and effective resistance.
\newblock \arxiv{1912.04196}, 2019.

\bibitem[Ric07a]{richter2007almost}
Peter~C. Richter.
\newblock Almost uniform sampling via quantum walks.
\newblock {\em New Journal of Physics}, 9(3):72, 2007.
\newblock \arxiv{quant-ph/0606202}.

\bibitem[Ric07b]{richter2007quantum}
Peter~C. Richter.
\newblock Quantum speedup of classical mixing processes.
\newblock {\em Physical Review A}, 76(4):042306, 2007.
\newblock \arxiv{quant-ph/0609204}.

\bibitem[Sze04]{szegedy2004quantum}
Mario Szegedy.
\newblock Quantum speed-up of {M}arkov chain based algorithms.
\newblock In {\em Proceedings of the 45th {IEEE} Symposium on Foundations of
  Computer Science ({FOCS})}, pages 32--41. IEEE, 2004.
\newblock \arxiv{quant-ph/0401053}.

\bibitem[Tao14]{taoblog}
Terence Tao.
\newblock Discretised wave equations.
\newblock {Available} at {\texttt{https:/terrytao.wordpress.}}\linebreak
  \texttt{com/2014/11/05/discretised-wave-equations/}, 2014.

\bibitem[Var85]{varopoulos1985long}
Nicholas Varopoulos.
\newblock {Long range estimates for Markov chains}.
\newblock {\em Bulletin des sciences math{\'e}matiques (Paris 1985)},
  109(3):225--252, 1985.

\bibitem[WA08]{wocjan2008speedup}
Pawel Wocjan and Anura Abeyesinghe.
\newblock Speedup via quantum sampling.
\newblock {\em Physical Review A}, 78(4):042336, 2008.
\newblock \arxiv{0804.4259}.

\end{thebibliography}

\appendix

\end{document}